\documentclass[12pt,preprint]{aastex}
\usepackage{epsfig}
\usepackage{multirow}
\usepackage{longtable}
\usepackage{natbib}
\usepackage{rotating}
\usepackage{amsmath, amsthm, amssymb, amsfonts}

\newcommand{\agile}{\textit{Agile}}
\newcommand{\rband}{$r'$--band}
\newcommand{\Iband}{$I$--band}

\newcommand{\mearth}{M$_\oplus$}

\newcommand{\tmcmc}{\texttt{TMCMC}}
\newcommand{\mtqlong}{\texttt{MultiTransitQuick}}
\newcommand{\mtq}{\texttt{MTQ}}

\newcommand{\boldtheta}{\boldsymbol{\theta}}

\newcommand{\iorb}{$i_{\text{orb}}$}

\newcommand{\eq}{Eq.}
\newcommand{\ie}{\textit{i.e.}}
\newcommand{\apostle}{APOSTLE}
\newcommand{\hst}{\textit{HST}}

\newcommand{\spitzer}{\textit{Spitzer}}
\newcommand{\keptel}{\textit{Kepler}}

\newcommand{\asec}{$"$}
\newcommand{\degree}{$^{\circ}$}
\newcommand{\xo}{XO-2}

\newcommand{\xob}{XO-2b}

\newcommand{\ldc}{LDC}
\newcommand{\sect}{$\S$}
\newcommand{\tresagilesect}{3}

\newcounter{parnum}
\newcommand{\N}{%
   \noindent\refstepcounter{parnum}%
    \makebox[\parindent][l]{\textbf{\Roman{parnum}.}}}
\begin{document}
\title{APOSTLE: Longterm Transit Monitoring and Stability Analysis of \xob}
\author{
  P.~Kundurthy\altaffilmark{1},
  R.~Barnes\altaffilmark{1,2},
  A.C.~Becker\altaffilmark{1},
  E.~Agol\altaffilmark{1},
  B.F.~Williams\altaffilmark{1},
  N.~Gorelick\altaffilmark{3},
  A.~Rose\altaffilmark{1}
}
\altaffiltext{1}{Astronomy Department, University of Washington, Seattle, WA 98195}
\altaffiltext{2}{Virtual Planetary Laboratory, USA}
\altaffiltext{3}{Google Inc., Mountain View, CA 94043}

\begin{abstract}
The Apache Point Survey of Transit Lightcurves of Exoplanets (APOSTLE) observed 10 transits of \xob\ over a period of three years. We present measurements which confirm previous estimates of system parameters like the normalized semi-major axis ($a/R_{\star}$), stellar density ($\rho_{\star}$), impact parameter ($b$) and orbital inclination ($i_{\text{orb}}$). Our errors on system parameters like $a/R_{\star}$ and $\rho_{\star}$ have improved by $\sim 40 \%$ compared to previous best ground-based measurements. Our study of the transit times show no evidence for transit timing variations and we are able to rule out co-planar companions with masses $\ge 0.20$\mearth\ in low order mean motion resonance with \xob. We also explored the stability of the \xo\ system given various orbital configurations of a hypothetical planet near the 2:1 mean motion resonance. We find that a wide range of orbits (including Earth-mass perturbers) are both dynamically stable and produce observable TTVs. We find that up to 51\% of our stable simulations show TTVs that are smaller than the typical transit timing errors ($\sim$20~sec) measured for \xob, and hence remain undetectable.
\end{abstract}

\section{Introduction} \label{sec_intro}
Observational efforts on two fronts (ground-based and space-based) have revealed a great diversity in the properties of planets (and planet candidates), and posed several new questions about the planet formation process. The search for planets and the careful measurement of planetary properties are the observational foundations upon which the physics of planet formation may be understood. The exoplanet community has co-opted the term `architecture' to embody several properties of planetary systems such as multiplicity and orbital parameters like eccentricities ($e$), inclinations (\iorb), semi-major axes ($a$) and orbital periods. The distributions of individual planetary properties like masses and radii ($M_{p}$ and $R_{p}$, respectively) may also be included in this term. The primary goal of the Apache Point Observatory Survey of Transit Lightcurves of Exoplanets (\apostle) is to catalog transit lightcurves at high precision, in order to (1) measure system parameters, and (2) look for transit timing variations that may indicate the presence of additional planets.

The transit technique applies to those systems where the orbital inclination of an exoplanet is close to $90$\degree\ (\ie\ edge-on) with respect to the observer's sky-plane \citep[see discovery paper][]{charbonneau00}. In this case, the observer sees a u-shaped dip in the starlight caused when the planet eclipses the star \citep{winn11}. The objective of several ground-based and space-based efforts focused on transit observations is to catalog and improve measurements of system parameters, which in-turn gives us an improved picture of the architecture of the planetary system. This process is key toward developing theories of planet formation that can adequately explain the origin and evolution of all planetary systems (including our own).

The \apostle\ target \xob\ is a Hot-Jupiter on a 2.6 day orbit around an early type K dwarf \citep[V = 11.2,][]{burke07}. The planet is known to have a mass of 0.555 $M_{\text{Jup}}$ and a radius of 0.992 $R_{\text{Jup}}$ \citep{southworth10}. The system is not known to have any other planet-mass objects. Transit timing measurements from the ground are consistent with a linear ephemeris \citep{fernandez09,sing11,crouzet12}, although the cited studies note that there are statistically significant deviations in the measurement of the orbital periods. A search for additional eclipses using the EPOXI mission by \citet{ballard11} had inconclusive results since the eclipse was not fully sampled. \spitzer\ observations of the secondary eclipse of \xob\ show IRAC 3.6, 4.5 and 5.8 $\mu m$ fluxes which are consistent with the presence of a temperature inversion \citep{machalek09}. There have also been detections of optical absorbers in the planetary atmosphere such as potassium from narrow band optical transmission spectrophotometry \citep{sing11}. Early theoretical studies indicated that stellar insolation levels directly influenced the presence or absence of a thermal inversion layer \citep[depending on the survival of atmospheric absorbers][]{hubeny03,burrows07,fortney08}. The planetary atmosphere classification system developed by \citet{fortney08} places \xob\ in an transition zone between planets with (pM) and without (pL) thermal inversions. \xob\ is one of a handful of Hot-Jupiters in this region. However, observational evidence suggests \citep{machalek08} that more complicated models need to be considered. Irradiation from stellar activity may need to be included \citep{knutson10} and atmospheric chemistry may also need to be considered to provide a more complete picture \citep{madhusudhan12}. The host star of the \xo\ system is known to be a high metallicity ([Fe/H] = 0.45 $\pm$ 0.02), high proper motion star ($\mu_{\text{Tot}}$ = 157 mas/yr) in a visual binary \citep{burke07}. Spectral activity indices show that \xo\ is fairly inactive compared to other stars of similar spectral type \citep{knutson10}.

A primary goal of APOSTLE and other campaigns that monitor transiting exoplanets is to search for TTVs that reveal the presence of unseen companions \citep{agol05,holmanmurray05}. In principle, Earth-mass planets in or near mean motion resonances could perturb the orbit of the transiting planet enough to produce a sinusoidal oscillation in the mid-points of transits. However, the full range of stable orbits that can produce a detectable TTV signal has never been explored. The detection of TTVs by Kepler \citep[e.g.][]{holman10} has demonstrated that stable systems are capable of producing TTVs, and other studies have explored a limited range of architectures \citep{haghighipourkirste11}, but the systematic exploration of parameter space of an Earth-mass planet in orbit near a hot Jupiter has not been undertaken. Here we examine 3.6 million possible masses and orbits of an approximately Earth-mass planet orbiting in or near the 2:1 outer mean motion resonance through N-body simulations. We find that stable orbits that also produce detectable TTVs do exist in the XO-2 system, and hence can exist in similar Hot-Jupiter systems.

In this paper we report observations of 10 transits of \xob, taken as part of \apostle. In \sect\ \ref{sec_obs} we outline our observations. In \sect\ \ref{sec_analysis} we briefly outline (i) the data reduction, photometry (\sect\ \ref{sec_reduction}), (ii) the transit model (\sect\ \ref{sec_mtq}) and lightcurve fitting (\sect\ \ref{sec_tmcmc}); both processes have also been described in previous work \citep{kundurthy13}. In \sect\ \ref{sec_sysparams} we present our estimates of the system parameters for \xob\ and in the subsections \sect\ \ref{sec_tdv} and \sect\ \ref{sec_ttv} we present results from our study of transit depth variations (TDVs) and transit timing variations (TTVs). In \sect\ \ref{sec_dynamics} we present results form N-body simulations used to study the stability of hypothetical planetary configurations at the 2:1 mean-motions resonance. Finally, in \sect\ \ref{sec_conclusions} we summarize our findings.

\section{Observations} \label{sec_obs}
\xob\ was observed by members of the \apostle\ team on 10 occasions over a timespan of 3 years from early 2008 until the spring of 2011. All observations of \xo\ were carried out using \agile, a high-speed frame-transfer CCD \citep{mukadam12}, on the ARC\footnote{Astrophysical Research Consortium} 3.5m telescope at Apache Point, New Mexico. The summary of observations is given in Table \ref{table_ObsSum_XO2}. \xo\ was an early \apostle\ target and was observed using a variety of instrumental settings, as the team had not converged on an optimal observing strategy prior to 2010. Early observations were made in the \Iband\ \citep[$\lambda_{0} = $ 805nm,][]{cousins76,bessell90} with several data sets taken with short read-out (exposure times, Column (5) in Table \ref{table_ObsSum_XO2}). The short read-out mode allowed for fine sampling of the lightcurve, but due to the lower signal-to-noise and the unsuitability for characterizing systematics, this observing mode was abandoned. In addition, the \Iband\ images also contained a strong contribution from a fringe pattern due interference from the backscattering of atmospheric lines within the CCD's pixels. The removal of this fringe pattern is discussed in \sect\ \ref{sec_reduction}. Early in 2010, the observing strategy changed to longer readouts (typically 45--75 seconds) to reduce the level of correlated noise (red-noise), and were made using the \rband\, which is similar to the SDSS\footnote{Sloan Digital Sky-Survey} $r$ filter \citep[$\lambda_{0} = $ 626nm,][]{fukugita96}, to reduce the influence of fringes. In the long read-out mode the telescope was defocused to spread the stellar point-spread function (PSF) across multiple pixels, which minimized the systematics caused by pixel-to-pixel wandering of the PSF over the imperfect flatfield. The longer exposures also allowed for a greater count rate which maximized the signal-to-noise per image. The count rate was kept below \agile's non-linearity limit of $\sim$52k ADU and well below its saturation level of 61k ADU by small adjustments to the telescope's secondary focus during observations.

\begin{sidewaystable}[!h]
\begin{center}
\caption{\label{table_ObsSum_XO2} APOSTLE Observing Summary for XO2}
\begin{tabular}{ccrccccrrcc}
\hline \hline
T\# & UTD & Obs. Cond. & Filter & Exp. & Bin & Phot. Ap. &RMS (ppm) & \%Rej. & Flux Norm. & Error Scaling \\
(1) & (2) & (3) & (4) & (5) & (6) & (7) &(8) & (9) & (10) & (11) \\
\hline
1&2008-01-09&Clear&I&0.5&45&13&557&3\%&0.9561&1.1740\\
2&2008-02-12&Clear&I&0.5&45&24&510&$ < 1\%$&0.9561&1.3129\\
3&2008-03-04&Clear&I&0.5&45&15&411&$ < 1\%$&0.9632&1.2510\\
4&2008-11-23&Poor Weather&I&10,25,45&45&27&939&1\%&0.9756&2.7482\\
5&2009-02-07&Poor Weather&I&45&-&31&1085&11\%&0.9721&2.9868\\
6&2009-03-13&Poor Weather&I&45&-&46&405&11\%&0.9766&1.2356\\
7&2010-10-25&Clear&r'&45&-&35&553&2\%&0.9676&2.2760\\
8&2010-12-27&Poor Weather&r'&45&-&43&775&5\%&0.9643&3.0092\\
9&2011-01-30&Clear&r'&45&-&48&354&1\%&0.9646&1.4571\\
10&2011-03-05&Clear&r'&45&-&43&693&$ < 1\%$&0.9660&2.7990\\
\hline
\end{tabular}
 \footnotesize \begin{tabular}{l}
(1) Transit Number, (2) Universal Time Date, (3) Observing Conditions, (4) Observing Filter, (5) Exposure Time (seconds) \\
(6) Bin size in seconds, (7) Optimal Aperture Radius (pixels), (8) Scatter in the residuals \\
(9) \% frames rejected due to saturation or other effects, (10) Flux normalization between the target and comparison star \\
(11) The factor by which the photometric errors were scaled \\
\end{tabular}
 \normalsize \end{center}
\end{sidewaystable}

\xo\ (TYC 3413-5-1) and its visual binary companion TYC 3413-210-1 (separated by $\sim$ 30\asec) were the only bright stars that fit in \agile's field of view. Both stars are of identical spectral type (K0V) and nearly identical brightnesses in the filters used by APOSTLE, with their Johnson R and I magnitudes at 10.8 and 10.5 respectively \citep{monet03}. Our uncalibrated differential photometry also showed good agreement in their brightnesses, with the out-of-eclipse, un-normalized flux ratios being different by only 4\% and 3\% in the \Iband\ and \rband, respectively (Column `Flux Norm.' in Table \ref{table_ObsSum_XO2}). The observations were made over a variety of observing conditions (Column `Obs. Conditions' in Table \ref{table_ObsSum_XO2}). The observing conditions are classified as `Clear' or `Poor Weather' with the former implying good data with few or no interruptions in data collection, and the latter indicating that the we experienced cloud cover or poor seeing conditions resulting in lower quality data. The tabulated transits are those for which we were able to capture the whole transit or at least a partial transit. Partial transits are those where portions of the in-eclipse lightcurve were lost due to bad weather or instrumental failure. Several data points were lost for the nights of UTD 2008-09-22 ($\# 4$), 2009-02-06 ($\# 5$), 2009-03-12 ($\# 6$) and 2010-12-27 ($\# 8$). However we do include these nights since we did manage to obtain reasonable portions of the in-eclipse and out-of-eclipse data, which make it possible to determine transit properties (albeit at a loss of accuracy and precision).

\section{Data Analysis} \label{sec_analysis}
This section outlines various stages in the analysis of lightcurves, starting with (i) the image reduction and photometry, and (ii) the transit model and Markov Chain Monte Carlo analyzer \citep[see also,][]{kundurthy13}.

\subsection{Reduction \& Photometry} \label{sec_reduction}
\apostle\ data were reduced using a pipeline developed specifically for \agile\ images. The pipeline (written in IDL\footnote{Interactive Data Language}) performs pixel-by-pixel error propagation, and image processing specific to the \agile\ CCD. In addition to the standard photometric reduction steps like dark subtraction and flat-fielding, the pipeline performs non-linearity and fringe corrections specific to \agile. The details on \agile's non-linearity correction are described in \sect\ \tresagilesect\ of \citet{kundurthy13}. Some of the initial observations by \apostle\ were carried out using short exposures (see Column (5) `Exp.') in Table \ref{table_ObsSum_XO2}. We binned these data by averaging the flux ratios in 45~sec bins.

For several of the initial \xo\ data taken using \agile's \Iband, a fringe pattern had to be subtracted to create science images. Photons from strong atmospheric lines (in the \Iband\ bandpass) backscatter within the CCD pixels, and owing to the variable thickness of the pixel array, interference between these photons creates the fringe pattern. Since the fringe pattern is convolved with the illumination pattern of the CCD, the fringe correction has to be applied after dark-subtraction and flat-fielding. During the initial characterization of the \agile\ CCD, an empirical fringe frame was produced by median combining dithered frames on the dark sky, where the flux contributions from stars were removed by outlier rejection. The resulting combined frame served as a ``Model'' fringe pattern ($F$), normalized to have a median of zero, and an amplitude of one, such that it could be scaled to match the fringe patterns on science frames. The science frame affected by the fringe pattern ($T^{'}$) is assumed to be a linear combination of the fringe-less science image ($T$) and a fringe pattern:
\begin{equation}
T^{'} = T + a_{0} F
\label{eq_fringe}
\end{equation}
where $F$ is the model fringe frame and $a_{0}$ is a scaling factor describing the amplitude of the fringing on a given frame. The fringe amplitude is estimated by minimizing $\chi^{2}$ and fitting for $a_{0}$ in the above model. The corrected images ($T = T^{'} - a_{0}F$), were found to be sufficiently corrected of fringes after visual examination. We found the fringe amplitudes ($a_{0}$) on \xo\ science frames to always be smaller than the standard deviation (\ie\ scatter) of the global sky background on each frame; $a_{0}$ ranged between 6--13$\%$ of the scatter in the sky for \xo\ data.

We extracted photometry from an optimal circular aperture centered around the target and comparison stars. In addition we extracted the counts on image products like the master-dark and master-flat, to serve as nuisance parameters for detrending, using the same aperture and centroids from photometry. The centroids were derived using SExtractor \citep{bertinarnouts96}, which allowed for the use of customized donut-shaped convolution kernels for defocused PSFs. Science frames where pixels inside a photometric aperture exceeded \agile's saturation limit of 61k were rejected. Images at the other extreme, where the stars were obscured by clouds, and resulted in low signal to noise measurements were also rejected (\ie\ where individual photometric errors were $>$ 5000~ppm). The fraction of rejected frames per night is listed in Column (9) `$\%$Rej.' in Table \ref{table_ObsSum_XO2}. The optimal aperture was selected after extracting photometry on a list of circular apertures with radii between 5--50 pixels at an interval of 1 pixel. The optimal aperture was selected where the scatter in the residuals of the detrended lightcurve minus a trial transit model (based on values from the literature) was minimized. The correction function $F_{cor}$ (or detrending function) is modeled as a linear sum of nuisance parameters as described by the following equation:
\begin{equation}
F_{cor,i} = \sum_{k=1}^{N_{\text{nus}}} c_{k} X_{k,i},
\label{eq_detrending}
\end{equation}
where $X_{k,i}$ are the nuisance parameters, $c_{k}$ are the corresponding coefficients. The index $k$ counts over the number of nuisance parameters $N_{\text{nus}}$. The detrending coefficients are chosen by minimizing the $\chi^2$ between the observed data ($O$), a model function ($M$) and the correction function,
\begin{equation}
\chi^2 = \sum_{j}^{N_{\text{all}}} \frac{(O_{j} - M_{j} - F_{\text{cor},j})^2}{\sigma_{j}^2}
\label{eq_chisq_multi}
\end{equation}
here $j$ is the index which counts over the total number of data points ($N_{\text{all}}$). The observed, model and correction function terms are all in normalized flux ratio units. The list of parameters used for detrending each lightcurve in the \xo\ dataset are presented in Table \ref{table_nuisance_XO2}, with variable definitions given in the footnotes. A suitable set of detrending parameters were selected for a given night by running several manual trials of linear least squares minimization on single transit lightcurves and its corresponding model lightcurve and correction function (\eq\ \ref{eq_detrending}). The model parameters were fixed at reasonable values and only the coefficients ($c_{k}$) were fit. Those nuisance parameters which returned large uncertainties on the coefficients were excluded since this indicated a poor match to the noise trend in a lightcurve. This method of selecting nuisance parameters is admittedly ad-hoc; automated parameter selection techniques could be applied in the future. The detrending is performed using the final set of nuisance parameters in conjunction with fitting the transit parameters in order to ensure that the correction process is not biased by the trial model used in the selection of nuisance parameters.

\begin{table}[!h]
\caption{\label{table_nuisance_XO2} Nuisance Parameters used for XO-2 Nights}
\begin{tabular}{ll}
\hline \hline
T\# & Nuisance Parameters used (FLDC, OLDC and MDFLDC) \\
\hline
T1&airmass, msky$_{1}$, msky$_2$, gsky, x$_1$, y$_1$, x$_2$, y$_2$, sD$_1$, sD$_2$, sF$_1$, sF$_2$, a$_0$\\
T2&airmass, msky$_{1}$, msky$_2$, gsky, x$_1$, y$_1$, x$_2$, y$_2$, sD$_1$, sD$_2$, sF$_1$, sF$_2$, a$_0$\\
T3&airmass, msky$_{1}$, msky$_2$, gsky, x$_1$, y$_1$, x$_2$, y$_2$, sD$_1$, sD$_2$, sF$_1$, sF$_2$, a$_0$\\
T4&airmass, msky$_{1}$, msky$_2$, gsky, x$_1$, y$_1$, x$_2$, y$_2$, sD$_1$, sD$_2$, sF$_1$, sF$_2$, a$_0$\\
T5&airmass, x$_1$, y$_1$, sD$_1$, sF$_1$, a$_0$\\
T6&airmass, x$_1$, y$_1$, sD$_1$, sF$_1$, a$_0$\\
T7&airmass, gsky, x$_1$, y$_1$, sD$_1$, sF$_1$\\
T8&airmass, gsky, x$_1$, y$_1$, sD$_1$, sF$_1$\\
T9&airmass, gsky, x$_1$, y$_1$, sD$_1$, sF$_1$\\
T10&airmass, gsky, x$_1$, y$_1$, sD$_1$, sF$_1$\\
\hline
\end{tabular}
\small

\begin{tabular}{l}
airmass = atmospheric column \\
(x$_{1}$, y$_1$), (x$_2$, y$_2$) = Centriods of Target (1) and Comparison (2)\\
msky$_1$, msky$_2$ = Median sky around Target (1) and Comparison (2) \\
gsky = Global median sky \\
sD$_1$, sD$_2$ = Sum of counts in aperture on Master-Dark \\
sF$_1$, sF$_2$ = Sum of counts in aperture on Master-Flat \\
a$_0$ = Fringe scaling (see Eq. \ref{eq_fringe}) \\
\end{tabular}
\end{table}

\begin{table}
\begin{center}
\caption{\label{table_LCdata_XO2} APOSTLE Lightcurve data* for XO-2}
\begin{tabular}{ccccc}
\hline \hline 
T\# & T-T0 & Norm. Fl. Ratio & Err. Norm. Fl. Ratio & Model Data \\
(1) & (2) & (3) & (4) & (5) \\
 \hline
1&-0.1341616&1.000004100&0.001207284&1.000000000\\
1&-0.1315574&0.999797978&0.001102731&1.000000000\\
1&-0.1305157&0.997707769&0.001047412&1.000000000\\
1&-0.1299949&0.999906661&0.001190794&1.000000000\\
1&-0.1289532&0.998693635&0.001118320&1.000000000\\
1&-0.1279116&0.998939727&0.001098241&1.000000000\\
1&-0.1273907&0.999021973&0.000987372&1.000000000\\
1&-0.1268699&1.001381660&0.001209313&1.000000000\\
1&-0.1263491&0.998899046&0.001076096&1.000000000\\
1&-0.1258282&1.001212353&0.001160014&1.000000000\\
 . & . & . & . & . \\
 . & . & . & . & . \\
\hline
\end{tabular}
\footnotesize \begin{tabular}{l} 
*The data are presented in their entirety as an online-only table. \\
(1) Transit Number, (2) Time Stamps - Mid Transit Times (BJD) \\
(3) Normalized Flux Ratio, (4) Error on Normalized Flux Ratio \\
(5) Model Data \\
 \end{tabular}
\end{center}
\end{table}

\clearpage
\begin{figure}[!h] 
\begin{center} 
\includegraphics[width=0.75\textwidth]{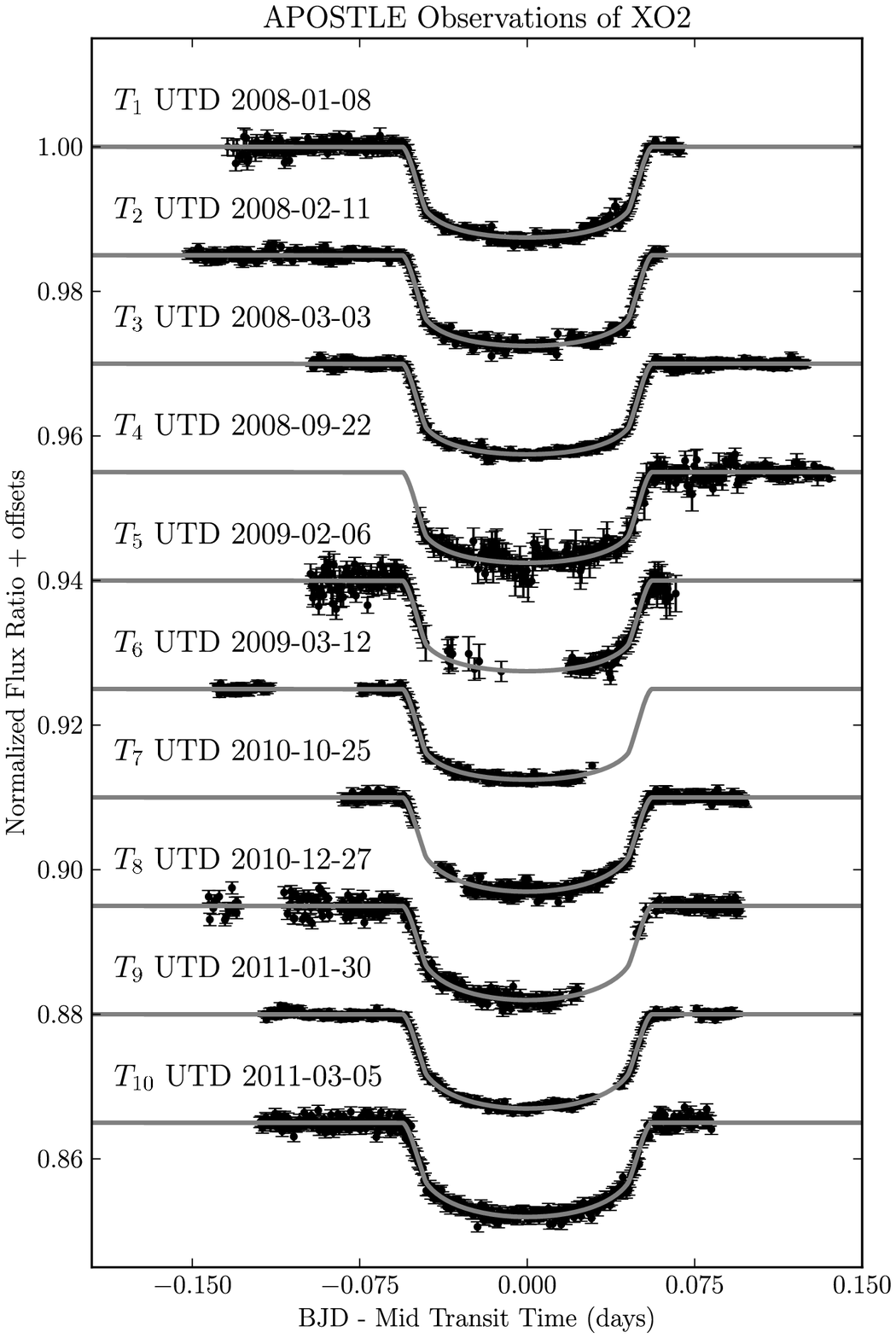}
\caption{\label{figure_LC_XO2}Six \Iband\ and four \rband\ lightcurves of \xob. The vertical axis is in normalized flux ratio units. The horizontal axis shows time from the mid-transit time in days, computed by subtracting the appropriate mid-transit time for each transit from the best-fit values in the Fixed LDC chain.}
\end{center} 
\end{figure} 

The 10 transits of \xob\ are shown in Figure \ref{figure_LC_XO2} in normalized flux ratios (with offsets for clarity). The plotted data result from the data reduction and model fitting processes described in sections \sect\ \ref{sec_reduction}, \sect\ \ref{sec_mtq} and \sect\ \ref{sec_tmcmc}.

\subsection{MultiTransitQuick} \label{sec_mtq}
We developed a transit model called \mtqlong\ (\mtq) in PYTHON, which is based on the analytic lightcurve models presented in \citet{mandelagol02}, and the PYTHON implementation of some of its functions \citep[from EXOFAST by][]{eastman12}. The description of \mtq\ used for this study is described in some detail in \citet{kundurthy13}. \mtqlong\ can be used to (1) fit for transit depths for data taken with multiple filters (Multi-Filter model), or (2) fit each transit depth individually (Multi-Depth).

The set of parameters used for Multi-Filter version of MTQ is $\boldsymbol{\theta}_{\text{Multi-Filter}}$ = \{$t_{T}$, $t_{G}$, $D_{j...N_F}$, $v_{1,j...N_F}$, $v_{2,j...N_F}$, $T_{i...N_T}$\}, where $t_T$ is transit duration, $t_G$ is the limb-crossing duration and $T_{i}$ are the mid-transit times. The filter dependent parameters include the transit depth $D_{j}$ and the limb-darkening parameters $v_{1,j}$ and $v_{2,j}$ described in \citet{kundurthy13}. The subscripts ${i...N_T}$ and ${j...N_F}$ are used to denote multiple transits ($N_T$) and multiple filters ($N_F$) respectively. The \xo\ data were gathered using the \Iband\ and \rband, where $N_F$ was 2 and the number of transits $N_T$ was 10.

The parameter set is only slightly different for the Multi-Depth version, with $\boldsymbol{\theta}_{\text{Multi-Depth}}$ = \{$t_{T}$, $t_{G}$, $D_{i...N_T}$, $v_{1,j...N_F}$, $v_{2,j...N_F}$, $T_{i...N_T}$\}; the difference being the transit depth is now fit for each transit separately instead of each filter. The filters are still tracked to ensure the use of the correct limb-darkening parameters for each lightcurve. A single lightcurve is set as the ``reference'' lightcurve and used to internally compute several orbital parameters. For the \xo\ data set we used \# 3 since it had few gaps, and had the best photometric precision among the \Iband\ lightcurves. We did not use an \rband\ lightcurve as a reference since this filter is more affected by limb-darkening when compared to the \Iband\ and hence estimates of the planet-to-star size ratio ($R_{p}/R_{\star}$) (used in the computation of orbital parameters) may be susceptible to degeneracies. Using the Multi-Depth model aids in understanding transit depth variations over multiple epochs.

\subsection{Transit MCMC} \label{sec_tmcmc}
We developed a Markov Chain Monte Carlo (MCMC) analyzer called Transit MCMC (\tmcmc), based on the Metropolis-Hastings (M-H) algorithm \citep{gelman03,tegmark04,ford05}, with an adaptive step-size modifier \citep{colliercameron07b}. MCMC routines quantify the uncertainty distributions of model parameters (given the data) using Bayes' theorem, by sampling parameter space such that samples from high probability regions (low $\chi^2$) are selected at a greater rate than those from low probability regions. The final ensemble of sampled points the MCMC routine typically represent the uncertainty distributions of the model parameters. One must note that adaptive MCMCs are generally not considered to be truly Markovian in nature and their results are valid only if adaptation diminishes with time \citep{robertsrosenthal09}; a property that our chains do display.

For \apostle\ data sets, we explored system parameters using three different kinds of chains. Two of these were based on the Multi-Filter parameter set $\boldtheta_{\text{Multi-Filter}}$ described in \sect\ \ref{sec_mtq} and third used the Multi-Depth parameter set ($\boldtheta_{\text{Multi-Depth}}$). The two Multi-Filter chains used Fixed Limb-Darkening Coefficients (\ldc) and Open Limb-Darkening Coefficients (OLDC). For the Fixed \ldc\ chains (FLDC), the coefficients were simply fixed to values tabulated for the appropriate observing filter \citep{claretbloemen11}. For the Open \ldc\ chains, the limb-darkening parameters $v_{1}$ and $v_{2}$, for both \Iband\ and \rband\ data, are allowed to float. The ability to constrain stellar limb-darkening requires high precision data, such as those collected using the \hst\ \citep{brown01,knutson07}. Previous attempts to fit for limb-darkening on \apostle\ data have resulted in Markov chains that failed to converge \citep{kundurthy11,kundurthy13}. The third type of Markov chain was run on the Multi-Depth parameter set $\boldtheta_{\text{Multi-Depth}}$ described in \sect\ \ref{sec_mtq}. \apostle\ lightcurves were gathered over a long time-baseline, and statistically significant depth variations seen in the data may help shed light on the various phenomena responsible for depth variations (see \sect\ \ref{sec_mtq}), or point to limitations in the data and model.

Several of the preliminary steps for executing a chain using \tmcmc\ are described in \citep{kundurthy13}. These steps include (1) setting bounds and (2) running short single-parameter exploratory chains to determine a set of suitable starting jump-sizes for the long Markov Chains. We ran long chains of $2\times10^{6}$ steps from two different starting locations for each model scenario: Fixed \ldc, Open \ldc\ and Multi-Depth/Fixed \ldc. After completion we (1) cropped the initial stages of these chains to remove the burn-in phase, where the chain is far from the best-fit region, and (2) we exclude the stage where the chain is far from the optimal acceptance rate of 23$ \pm $ 5$\%$ \citep[as noted for multi-parameter chains,][]{gelman03}. We run three types of post-processing on the chains after cropping: (a) We compute the ranked and unranked correlations in the chains of every fit parameter with respect to the others. These statistics provide an estimate of the level of degeneracy between parameters in a given model. The next post-processing steps are two commonly used diagnostics to check for chain convergence, namely (b) computing the auto-correlation lengths and (c) the Gelman-Rubin \^{R}-static values \citep{gelmanrubin92}. The auto-correlation lengths determine the scale over which a chain has local trends. From the auto-correlation length one can compute the effective length as the total chain length divided by the auto-correlation length, which represents the statistical significance with which the uncertainty distribution was sampled. A large effective length ($> 1000$) represents a well-sampled distribution. The \^{R}-statistic represents the level of coverage the chain has over the parameter space. When parameter space has been properly sampled the \^{R}-statistic computed using chains from different starting locations will be close to 1. We deem those chains as coverged that have an effective length $>$ 1000 and an \^{R}-statistic within 10$\%$ of 1.

\subsubsection{TAP}
The Transit Analysis Package \citep[TAP;][]{gazak11} implements the red-noise model of \citet{carterwinn09}, who find that models that do not fit for red-noise are subject to inaccuracies in transit parameters on the order of 2-3$\sigma$ and tend to have underestimated errors by up to 30$\%$. For transit timing studies, poor estimates such as these are cause for concern, since smaller errors and large deviations from the expected time can easily lead to false claims of TTVs. Since \tmcmc\ does not include red-noise analysis we run fits on \apostle\ lightcurves using TAP as a check to the results derived from \tmcmc.

The typical TAP parameter set is: $\boldsymbol{\theta}_{\text{TAP}}$ = \{$a/R_{\star}$, $i_{\text{orb}}$, $(R_{p}/R_{\star})_{i...N_F}$, $T_{i...N_T}$, $\sigma_{(\text{white},i...N_T)}$, $\sigma_{(\text{red},i...N_T)}$\}, where $a/R_{\star}$, $i_{\text{orb}}$ and $(R_{p}/R_{\star})$ are the commonly fit transit parameters denoting the semi-major axis (in stellar radius units), the orbital inclination and the planet-to-star radius ratio, respectively. The noise analysis parameters $\sigma_{(\text{white},i...N_T)}$ and $\sigma_{(\text{red},i...N_T)}$ are the white-noise and red-noise levels for $N_T$ transits respectively. The TAP package does not fit for the period using the transit times, and often yields poor estimates of the period, so we fixed the period to the value derived from \tmcmc. The limb-darkening was fixed to values from the literature. The orbital eccentricity ($e$) and argument of periastron were kept fixed at 0 for \xob.

\section{System Parameters} \label{sec_sysparams}
This section describes results from our execution of the 2 chains for the parameter set $\boldtheta_{\text{Multi-Filter}}$, and 1 for the $\boldtheta_{\text{Multi-Depth}}$ described in \sect\ \ref{sec_tmcmc}. Post processing statistics and other data for these chains are listed in Table \ref{table_ChainStats_XO2}. The columns `$N_{\text{free}}$', `Chain Length', `Corr. Length' and `Eff. Length' list the number of free parameters, the length of the cropped chain, the correlation and effective lengths, respectively. All chains were run for approximately 2 Million steps, but about 100,000 of the initial steps were removed to account for ``burn-in'' and selection rate stabilization. The \xob\ chain with Open \ldc\ has a low effective length indicating poor Markov chain statistics. The Fixed LDC chains (both $\boldtheta_{\text{Multi-Filter}}$ and $\boldtheta_{\text{Multi-Depth}}$) model satisfy the condition of a well sampled posterior distribution (effective length is $> 1000$). The final two columns list the goodness of fit (\ie\ lowest $\chi^2$ in the MCMC ensemble) and Degrees-of-freedom (DOF) from the respective chain. Parameters from all chains had Gelman-Rubin \^{R}-statistics close to 1 indicating that the parameter space was covered evenly (though the OLDC chain was not sampled finely enough, based on the auto-correlation data).

\begin{table}[!h]\footnotesize
\begin{center}
\caption{\label{table_ChainStats_XO2} \tmcmc\ Chains for XO-2}
\begin{tabular}{cccccccc}
\hline \hline
Chain & Model Vector & N$_{\text{free}}$ &Chain Length & Corr. Length & Eff Length & $\chi^{2}$ & DOF \\
\hline
FLDC&$\boldtheta_{\text{Multi-Filter}}$&14&1,900,001&100&19,000&3272.94&3312\\
OLDC&$\boldtheta_{\text{Multi-Filter}}$&18&1,900,001&4,949&383&3268.48&3308\\
MDFLDC&$\boldtheta_{\text{Multi-Depth}}$&22&1,900,001&588&3,231&3197.53&3304\\
\hline
\end{tabular}
\end{center}
\end{table}

\begin{table}[!h]
\small
\begin{center}
\caption{\label{table_ParTable01_XO2} XO-2 Parameters for $\boldtheta_{\text{Multi-Filter}}$ }
\begin{tabular}{cccc}
\hline \hline
Parameter & FLDC & OLDC & Unit\\
\hline
\multicolumn{4}{c}{MTQ Parameters} \\
\hline
$t_{G}$&0.0107$\pm$0.0002&0.0108$\pm$0.0002&days\\
$t_{T}$&0.1008$\pm$0.0001&0.1004$\pm$0.0002&days\\
$D_{\textrm{(I)}}$&0.0126$\pm$0.0001&0.0126$\pm$0.0001&-\\
$D_{\textrm{(r')}}$&0.0130$\pm$0.0001&0.0131$\pm$0.0001&-\\
$v_1{\textrm{(I)}}$&(0.5944)&0.5440$\pm$0.0313&-\\
$v_1{\textrm{(r')}}$&(0.6980)&0.6411$^{+0.0332}_{-0.0249}$&-\\
$v_2{\textrm{(I)}}$&(0.1452)&0.2806$\pm$0.0904&-\\
$v_2{\textrm{(r')}}$&(0.3524)&0.4786$^{+0.0734}_{-0.0992}$&-\\
\hline
\multicolumn{4}{c}{Derived Parameters} \\
\hline
$(R_{p}/R_{\star})_{\textrm{(I)}}$&0.1030$\pm$0.0003&0.1033$\pm$0.0004&-\\
$(R_{p}/R_{\star})_{\textrm{(r')}}$&0.1024$\pm$0.0003&0.1029$\pm$0.0004&-\\
b&0.17$^{+0.04}_{-0.02}$&0.21$\pm$0.03&-\\
$a/R_{\star}$&8.14$\pm$0.06&8.11$\pm$0.07&-\\
$i_{orb}$&88.79$\pm$0.15&88.53$^{+0.02}_{-0.10}$&$^{o} (deg)$\\
$\nu/R_{\star}$&19.55$\pm$0.15&19.48$\pm$0.17&days$^{-1}$\\
$\rho_{\star}$&1.49$\pm$0.03&1.48$\pm$0.04&g/cc\\
P (2.6159 days +)&-3467$\pm$22&-3466$\pm$22&milli-sec\\
\hline
\end{tabular}
\end{center}
\end{table}

\begin{table}[!h]
\small
\begin{center}
\caption{\label{table_ParTable02_XO2} XO-2 Parameters for $\boldtheta_{\text{Multi-Depth}}$ }
\begin{tabular}{cccccc}
\hline \hline
Transit Depths & Value & Units & $R_{p}/R_{\star}$ & Value & Units \\
\hline
 & & \Iband\ & & & \\
 \hline
$(D)_{1}$&0.01338$\pm$0.00013&-&$(R_{p}/R_{\star})_{1}$&0.1061$\pm$0.0005&-\\
$(D)_{2}$&0.01300$\pm$0.00017&-&$(R_{p}/R_{\star})_{2}$&0.1047$\pm$0.0007&-\\
$(D)_{3}$&0.01226$\pm$0.00013&-&$(R_{p}/R_{\star})_{3}$&0.1019$\pm$0.0005&-\\
$(D)_{4}$&0.01199$\pm$0.00013&-&$(R_{p}/R_{\star})_{4}$&0.1009$\pm$0.0005&-\\
$(D)_{5}$&0.01209$\pm$0.00039&-&$(R_{p}/R_{\star})_{5}$&0.1013$\pm$0.0015&-\\
$(D)_{6}$&0.01267$\pm$0.00015&-&$(R_{p}/R_{\star})_{6}$&0.1035$\pm$0.0006&-\\
 \hline
 & & \rband\ & & & \\
 \hline
$(D)_{7}$&0.01326$\pm$0.00009&-&$(R_{p}/R_{\star})_{7}$&0.1032$\pm$0.0004&-\\
$(D)_{8}$&0.01320$\pm$0.00015&-&$(R_{p}/R_{\star})_{8}$&0.1030$\pm$0.0006&-\\
$(D)_{9}$&0.01307$\pm$0.00009&-&$(R_{p}/R_{\star})_{9}$&0.1026$\pm$0.0003&-\\
$(D)_{10}$&0.01300$\pm$0.00013&-&$(R_{p}/R_{\star})_{10}$&0.1024$\pm$0.0005&-\\
\hline
\multicolumn{6}{c}{Other MTQ Parameters} \\
\hline
Parameter & Value & Units & Parameter & Value & Units \\
\hline
$t_{G}$&0.0108$\pm$0.0001&days&$t_{T}$&0.1008$\pm$0.0001&days\\
$v_1{\textrm{(I)}}$&(0.5944)&-&$v_1{\textrm{(r')}}$&(0.6980)&-\\
$v_2{\textrm{(I)}}$&(0.1452)&-&$v_2{\textrm{(r')}}$&(0.3524)&-\\
\hline
\multicolumn{6}{c}{Derived Parameters} \\
\hline
b&0.24$\pm$0.02&-&$a/R_{\star}$&8.02$^{+0.03}_{-0.04}$&-\\
$i_{orb}$&88.29$\pm$0.15&$^{o} (deg)$&$\nu/R_{\star}$&19.27$^{+0.08}_{-0.10}$&days$^{-1}$\\
$\rho_{\star}$&1.43$\pm$0.02&g/cc&-&-&-\\
\hline
\end{tabular}
\end{center}
\end{table}

\begin{figure}[!h] 
\begin{center} 
\includegraphics[width=0.95\textwidth]{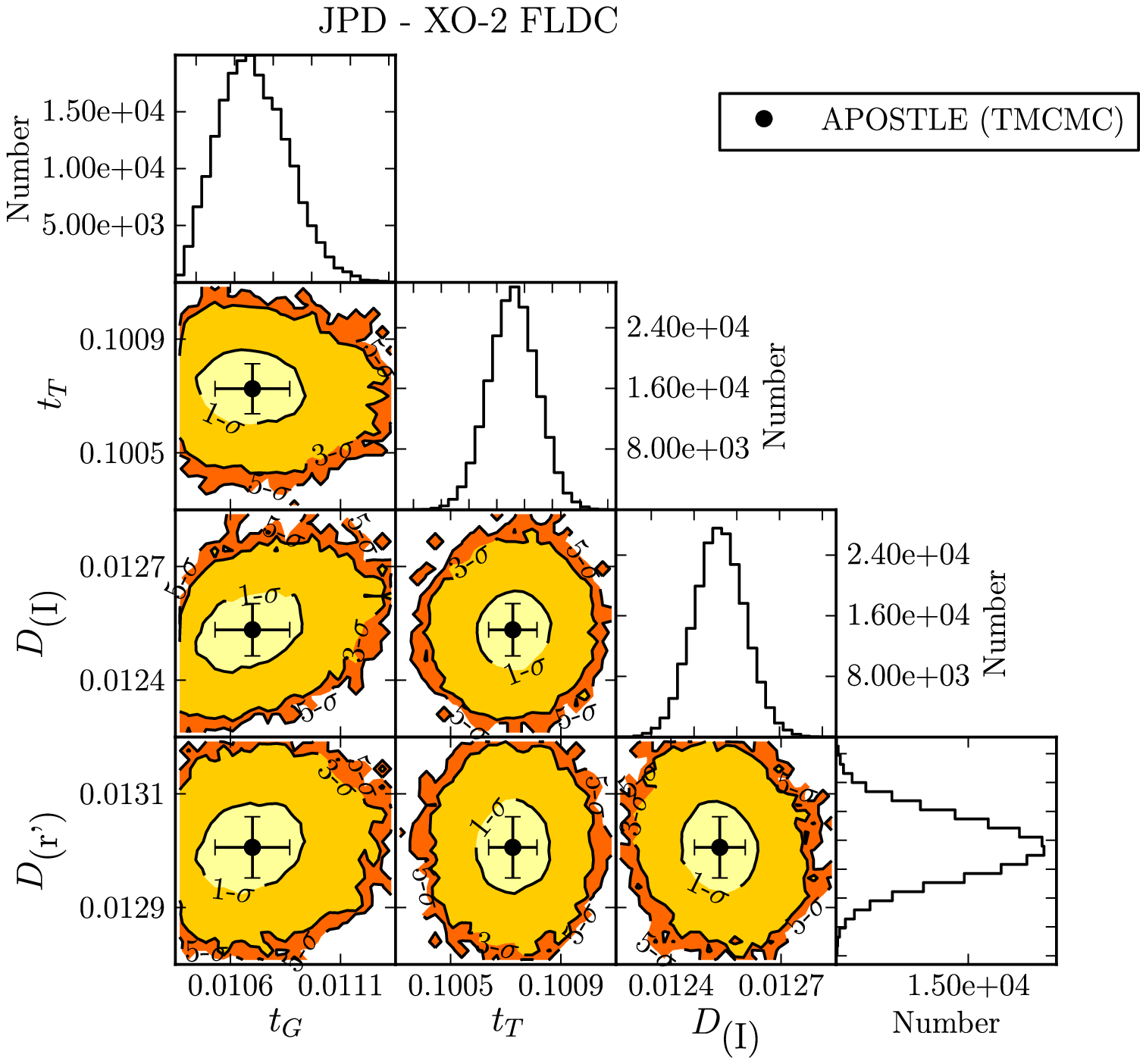}
\caption{\label{figure_MCMC_FLDC_XO2}Plots of the Joint Probability Distributions (JPD) of parameters from the Fixed LDC chains, demonstrating that parameters chosen for $\boldtheta_{\text{Multi-Filter}}$ are generally uncorrelated. Table \ref{table_ParTable01_XO2} gives units.}
\end{center} 
\end{figure}

\begin{figure}[!h] 
\begin{center} 
\includegraphics[width=0.95\textwidth]{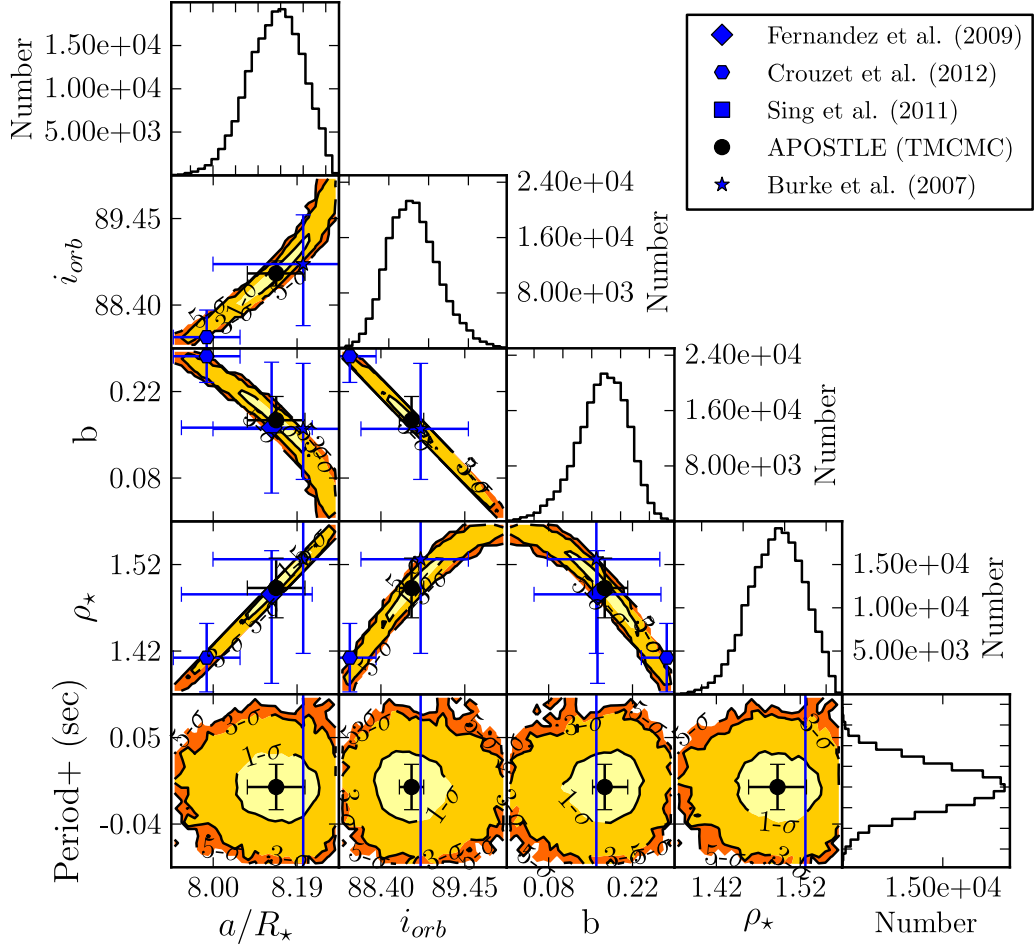}
\caption{\label{figure_MCMC_FLDC_DER_XO2}Plots of the Joint Probability Distributions (JPD) of derived system parameters from the Fixed LDC chains. Parameter estimates available in the literature are overplotted. Table \ref{table_ParTable01_XO2} gives units. }
\end{center} 
\end{figure}

The resulting best-fit parameter estimates are listed in Table \ref{table_ParTable01_XO2} for the Multi-Filter models, and in Table \ref{table_ParTable02_XO2} for the Multi-Depth models. These tables also list the derived system parameters. The transformation between the \mtq\ parameters to the derived system parameters are described in \citet{carter08} and \citet{kundurthy11}. Contour plots showing the joint probability distributions (JPDs) for the fit and derived parameters are shown in Figures \ref{figure_MCMC_FLDC_XO2} and \ref{figure_MCMC_FLDC_DER_XO2}, respectively. There are no strong correlations between the fit parameters $t_{T}$, $t_{G}$ and the transit depths, $D_{I}$ and $D_{r'}$, as seen in Figure \ref{figure_MCMC_FLDC_XO2}.

\clearpage

\clearpage

System parameters agree with previously published values in the literature, as seen by the overlap of the uncertainties in the JPD plot (Figure \ref{figure_MCMC_FLDC_DER_XO2}). In addition \apostle's measurements give tighter constraints on several of the system parameters. The errors on $a/R_{\star}$, $i_{\text{orb}}$, the impact parameter ($b$) and the stellar density ($\rho_{\star}$) are more precise than previous measurements by a factor of $\sim$3 \citep{burke07,fernandez09,sing11}. However, previous studies using the combination of the \tmcmc\ Markov chain analyzer and \mtq\ transit model have resulted in underestimated errors since we do not include rednoise analysis \citep{carterwinn09}. Hence, the errors presented in Table \ref{table_ParTable02_XO2} and Figure \ref{figure_MCMC_FLDC_DER_XO2} do not truly reflect improvements in the measurements of parameters. More conservative constraints were placed on a subset of these system parameters using the TAP package. Comparisons of some parameters and their uncertainties are presented in \ref{table_ParTable03_XO2}.

\begin{table}[!h]
\small
\begin{center}
\caption{\label{table_ParTable03_XO2} Comparison of Estimates of System Parameters for XO-2b}
\begin{tabular}{cccccccc}
\hline \hline \scriptsize 
Parameter & TMCMC & TAP & B07 & F09 & S11 & C12 & Units \\
\hline
$a/R_{\star}$ &8.14$\pm$0.06 & 8.13$\pm$0.10 & 8.20$\pm$0.20 & 8.13$\pm$0.20 & 7.83$\pm$0.17 & 7.99$\pm$0.07 & - \\
$i_{orb}$ &88.79$\pm$0.15 & 88.80$\pm$0.61 & 88.90$\pm$0.75 & - & 87.62$\pm$0.51 & 88.01$\pm$0.33 & $^{o} (deg)$ \\
b &0.172$^{+0.040}_{-0.021}$ & 0.171$\pm$0.085 & 0.158$\pm$0.110 & 0.160$\pm$0.110 & 0.324$\pm$0.070 & 0.280$\pm$0.044 & - \\
$\rho_{\star}$ &1.49$\pm$0.03 & 1.49$\pm$0.05 & 1.52$\pm$0.11 & 1.48$\pm$0.10 & 1.33$\pm$0.09 & 1.41$\pm$0.04 & g/cc \\
\hline
\end{tabular}
\footnotesize \begin{tabular}{l}
 TMCMC \& TAP values are from independent analysis of APOSTLE lightcurves \\
 B07 - \citep{burke07}, F09 - \citep{fernandez09}, S11 - \citep{sing11}, C12 - \citep{crouzet12} \\
\end{tabular}\end{center}
\end{table}
It is clear that using TAP on the \apostle\ dataset and accounting for rednoise provides more conservative estimates of the system parameters when compared to \tmcmc\ values. TAP errors for $a/R_{\star}$ and $\rho_{\star}$ are better than those reported by \citet{sing11} by upto $\sim$40\%; whose observations were made using the 10.4m Gran Telescopio Canarias (GTC). One must note the \citet{sing11} data result from narrow-band photometry and hence would have much lower photometric precision when compared to broadband observations from the same telescope. The resulting TAP errors for the impact parameter ($b$) and the orbital inclination are larger than those from the GTC study by $\sim$ 20\%. Thus we report only some improvements in the measurements of system parameters. Our system parameters agree with the current best estimates reported in \citet{crouzet12}, but our error bars are larger by factors $\leq 2$.

\subsection{Transit Depth Analysis} \label{sec_tdv}
\begin{figure}[!h] 
\begin{center} 
\includegraphics[width=0.95\textwidth]{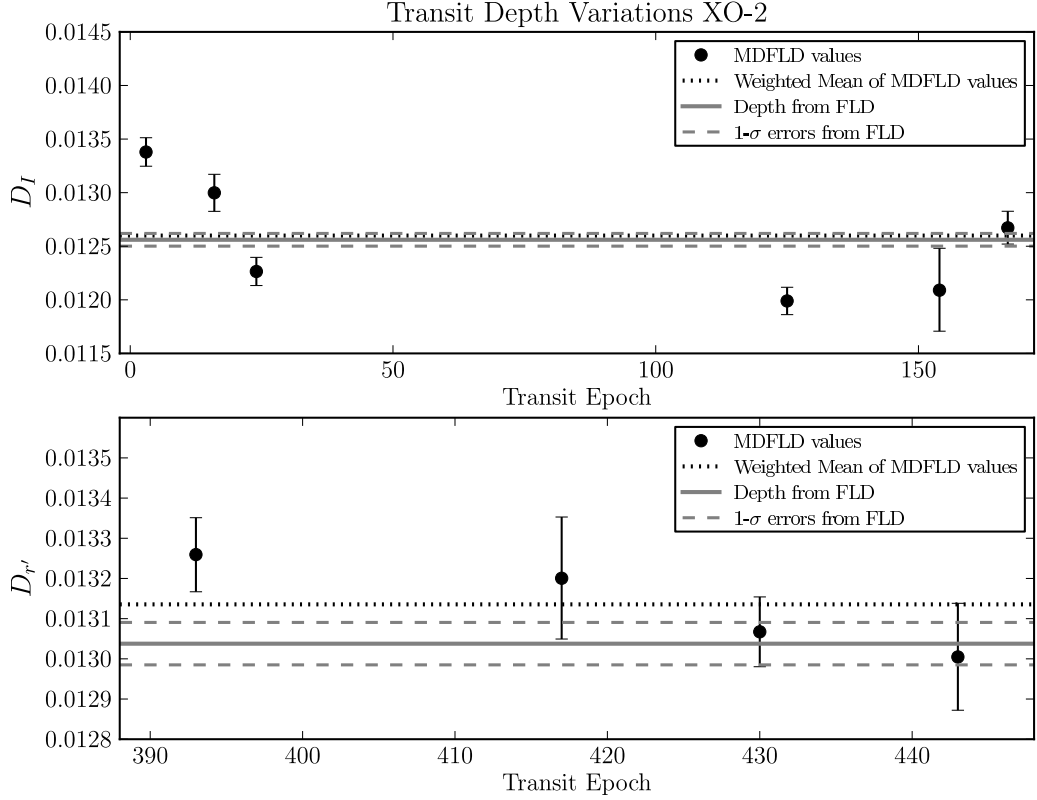}
\caption{\label{figure_TDepthV_XO2}The transit depth $D$ as a function of transit epoch for both \Iband\ and \rband\ observations of \xob. The solid horizontal and dashed lines represent the best-fit value and errors respectively for $D$ from the Fixed LDC \tmcmc\ fit. The dotted line is the weighted mean of transit depth values from the Multi-Depth Fixed LDC chains.}
\end{center} 
\end{figure} 

Figure \ref{figure_TDepthV_XO2} shows transit depth vs. transit epoch for the \Iband\ (top panel) and \rband\ (bottom panel). The overall variations in the \Iband\ depth are $\sim$ 0.05\% compared to the 0.01\% uncertainty in D$_{(I)}$ from the joint fit to depths in the Fixed \ldc\ chain. Depth variations can be caused by spots. Even though the variations we present are significant we refrain from claiming the detection of spot-modulation. These deviations are likely due to the incomplete sampling of several transits. Spots influence stellar brightness to a greater extent at shorter wavelengths, so the \rband\ would be more conducive to showing depth variations. However, the overall depth variation in the \rband\ lightcurves is of the order 0.01\% and is consistent with the errors on D$_{(r')}$ from the fit reported by the Fixed \ldc\ chain (Table \ref{table_ParTable01_XO2}). The variations seen in the \Iband\ depths are difficult to exaplin. They can either be due to (1) real brightness variations caused by spot-modulations, or (2) variations arising from the transit model's inability to accurately constrain transit depths and errors given the incomplete sampling of lightcurves (like Transit \# 4, 5 and 6). We note that errors on the transit depth from our Multi-Depth \ldc\ chains are more sensitive to the incompleteness of in-eclipse data rather than the photometric precision of a given lightcurve. For example, transit numbers \#4 and \#6 have 86\% and 72\% of the in-eclipse transit data sampled respectively; they also have lightcurve residuals of 939ppm and 405ppm respectively. So even though \# 6 has significantly better photometric precision, the depth for \# 4 is constrained slightly better due to the fact that more in-eclipse points are available. A more dramatic difference can be seen with transit \#5, which has only 47\% of the in-eclipse lightcurve sampled and the worst depth constraint. Other than \#5 all transits have eclipses sampled to better than 70\%, hence the depth uncertainties are all comparable. Even though the Multi-Depth LDC chain satisified the convergence criteria described in \S\ \ref{sec_tmcmc}, the effective length of the chain is far lesser than the value derived for the Multi-Filter \ldc\ chain (see Table \ref{table_ChainStats_XO2}), indicating the lower significance of these MCMC results. In addition one must note the lack of rednoise analysis in \tmcmc, which implies that the errors presented Figure \ref{figure_TDepthV_XO2} are probably underestimates.

\subsection{Transit Timing Analysis} \label{sec_ttv}
Several planetary systems have been observed to have Transit Timing Variations (TTVs) \citep{holman10,lissauer11a,ballard11,nesvorny12}. In certain configurations these variations can be on the order of minutes and can be easily seen in an diagram showing observed minus computed transit times (\ie\ the difference between the measured times and times expected from purely Keplerian orbital periods). Using TTVs to look for additional planets was first proposed by \citet{agol05} and \citet{holmanmurray05}, who showed that unseen planetary siblings can gravitationally influence the orbits of a known transiting planet. The TTV signals are known to be especially strong if the unseen companion lies close to mean motion resonance with the transiting planet.

Available transit times from the literature for \xob\ include those of the discovery paper by \citet{burke07}, and the follow-up observations by \citet{fernandez09}, \citet{sing11} and \citet{crouzet12}. We excluded the 11 transit times from \citet{burke07} since the reported timing errors were on the order of $\sim 3$min, and are too large to provide meaningful constraints on the orbital ephemeris. We do include transit times from the other three studies. The time coordinate BJD (TDB) has become the standard system used for transit timing studies \citep{eastman10} and all transit times used in this study were brought to this system. The timestamps of all \apostle\ data were converted to BJD (TDB) in the customized reduction pipeline \citep{kundurthy11}. The \apostle\ pipeline's time conversions have been verified by comparison to the commonly used time conversion routines made available by \citet{eastman10}.

\begin{table}[!h]
\begin{center}
\caption{\label{table_OC_XO2} APOSTLE Transit Times for XO2}
\begin{tabular}{ccccc}
\hline \hline
Epoch & T0 (\tmcmc) & $\sigma_{T0}$ & T0 (TAP) & $\sigma_{T0}$ \\
 & 2,400,000+ (BJD) & (BJD) & 2,400,000+ (BJD) & (BJD) \\
\hline
26&54474.73242&0.00011&54474.73252&0.00021\\ 
39&54508.73875&0.00013&54508.73877&0.00020\\ 
47&54529.66596&0.00009&54529.66604&0.00016\\ 
148&54793.86845&0.00024&54793.86836&0.00041\\ 
177&54869.72753&0.00023&54869.72735&0.00050\\ 
190&54903.73397&0.00014&54903.73403&0.00020\\ 
416&55494.91796&0.00015&55494.91794&0.00021\\ 
440&55557.69891&0.00020&55557.69892&0.00025\\ 
453&55591.70507&0.00008&55591.70511&0.00015\\ 
466&55625.71093&0.00014&55625.71112&0.00019\\ 

\hline 
\end{tabular}
\begin{tabular}{ccccc}
Fit & Period (days) & $\sigma_{\text{P}}$ & T0 (BJD) & $\sigma_{\text{T0}}$ \\
 \hline
\tmcmc & 2.615860095 & $\pm$ 0.000000209 & 2454474.7327333 & $\pm$ 0.0000599 \\
TAP & 2.615860014 & $\pm$ 0.000000346 & 2454474.7327964 & $\pm$ 0.0001028 \\
\hline
\end{tabular}
\end{center}
\end{table}

\begin{figure}[!h] 
\begin{center} 
\includegraphics[width=0.95\textwidth]{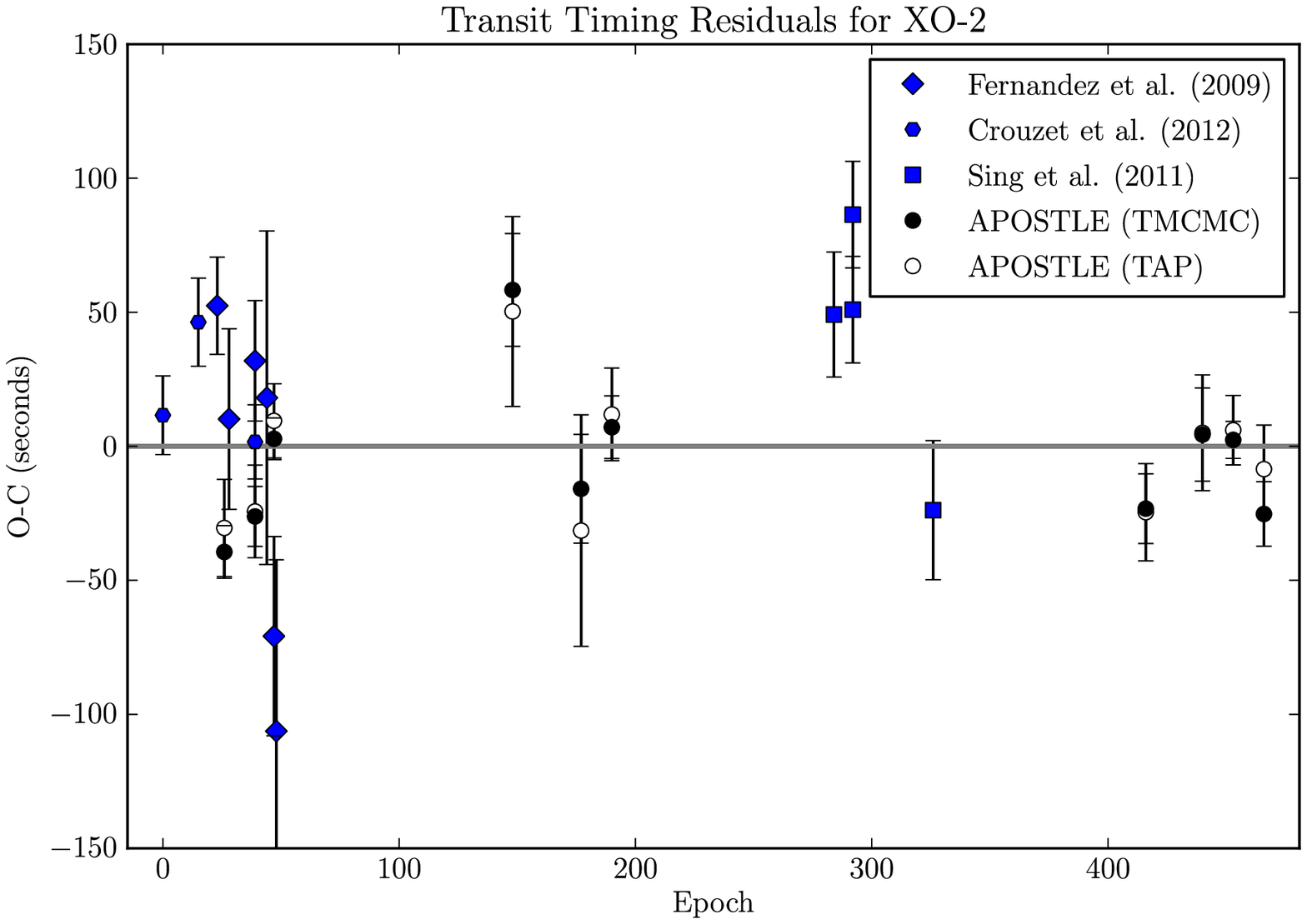}
\caption{\label{figure_OC_XO2}The Observed minus Computed Transit Times for \xob. Values from APOSTLE's \tmcmc\ fit, TAP and the literature are plotted. The horizontal axis represents the transit Epoch. The zero-line ephemeris is described in $\S$ \ref{sec_ttv}}
\end{center} 
\end{figure}

The observed minus computed mid-transit times (O-C) are plotted in Figure \ref{figure_OC_XO2}. There are two versions of APOSTLE transit times presented, one from the Fixed \ldc\ chain (\tmcmc\ + \mtq), and the other from the TAP fit to APOSTLE lightcurves. These transit times are tabulated in Table \ref{table_OC_XO2}. A linear ephemeris was fit to all the data (including literature values) using the equation,
\begin{equation}
T_i = T0 + \text{Epoch}_i \times P
\label{eq_transitephem}
\end{equation}
resulting in a best fit ephemeris of,
\begin{eqnarray*}
P = 2.61585988 \pm 0.00000016 & \text{days} \\
T0= 2454406.720516 \pm 0.000046 & \text{BJD} \\
\end{eqnarray*}
with a goodness of fit $\chi^{2} =$ 105.14 for 31 Degrees-of-freedom. The reduced chi-squared $\chi^{2}/\nu =$ 3.39 indicates that the linear fit is not robust, and is either due to the large timing deviations in the data or underestimated errors. For example, several of the \citet{sing11} data points lie far from the zero O-C line, with the largest deviation being $\sim 106$ sec. The scatter in the O-C values as a whole is $\sim 39$ sec which makes the $106$ sec deviation fall within the 3$\sigma$ confidence interval of the collective data set. Though the linear ephemeris does not precisely fit the timing data, the level of variation is not significant enough for us to claim unseen planets as the cause.

In order to compare the ephemerides derived with and without rednoise analysis, we fit for a linear ephemeris to the APOSTLE transit times from \tmcmc\ and TAP respectively (presented in Table \ref{table_OC_XO2}). The difference between the periods derived for these subsets and the period derived from all available transit times was $< 7$ milli-seconds. The results from the fits to the \tmcmc\ and TAP transit times are presented at the bottom of Table \ref{table_OC_XO2}. The reduced $\chi^{2}$s for linear ephemeris fits to the \tmcmc\ and TAP subsets were 3.96 and 1.26 respectively, confirming that TAP gives more conservative errors for the transit times thanks to the red-noise analysis.

We were also able to rule out sinusoidal trends in the data by running a generalized lomb-scargle analysis on the O-C data \citep{zechmeisterkurster09}. We fit for a period, amplitude and phase offset on two sets of O-C data. The first set included the literature dataset and all APOSTLE measurements, and the second included only the TAP measurements of APOSTLE. We found that a sinusoid of period $\sim$ 19 days and amplitude of $\sim$ 32 seconds improved the fit when the entire timing data set was used. The sinusoid fit yielded a $\chi^2 = $81.8, which was an improvement of $\Delta\chi^2 =$ 23.3 compared to the linear ephemeris (see above). However, the reduced $\chi^2$ remains greater than 1, making it a non-robust fit. The periodogram analysis on only the TAP data, yielded a period and amplitude of 8.5 days and 27.4 seconds respectively. The $\Delta \chi^2$ showed improvement compared to the linear ephemeris model by $\sim $7.6. In a manner similar to \citet{becker13} we tested the significance of this sinusoidal fit by repeating the analysis on 10$^5$ randomly cycled permutations of the amplitudes of the O-C measurements, keeping the epochs fixed. We found that 76.7\% of the fits showed improved $\Delta \chi^2$ at amplitudes greater than or equal to 27.4 seconds, indicating that the periodicity is not likely from a real TTV.

For the case when planets are in mean-motion resonance (MMR), \citet{agol05} showed that the analytic expression, $\delta t_{\text{max}} \sim \frac{P}{4.5 j} \frac{m_{\text{pert}}}{(m_{\text{pert}}+m_{\text{trans}})}$, can roughly estimate the amplitude of the timing deviation ($\delta t_{\text{max}}$). The quantities $m_{\text{pert}}$, $m_{\text{trans}}$, $P$ and $j$ are the mass of the unseen perturber, the mass of the transiting planet, the orbital period of the transiting planet and the order of the resonance respectively. For the \xo\ system, we can rule out possible system configurations given the amplitude of the weak sinusoidal fit to the TTV data ($\sim 27.4$ sec). Using the orbital period from Table \ref{table_OC_XO2} and the mass of \xob, $m_{\text{trans}}$ = 0.565 $M_{\text{Jup}}$ \citep{fernandez09}, we compute the maximum mass perturber that could exist in the \xo\ system in the 2:1 MMR to be $\sim 0.2 M_{\earth}$, \ie\ additional planets with $M_{p} < 0.2$ \mearth\ may exist near the 2:1 MMR given these data. At higher order resonances, this maximum mass (for a possible perturber) is larger. A more detailed analysis of TTV needs to account for a variety of orbital configurations of hypothetical companions, which is addressed in the following section.

\section{System Dynamics} \label{sec_dynamics}
In order to evaluate the stability of systems that produce a detectable TTV, we integrated $3.6 \times 10^6$ orbital configurations designed to mimic the \xo\ system. Each trial consisted of the known planet \xob\ and a hypothetical terrestrial-like companion.

The parameter space we cover is presented in Table \ref{table_stability_XO2}. In this table $\Delta$ is the interval between values, all of which were varied uniformly in the range between ``Min'' and ``Max,'' producing $N$ bins. In this table, $m$ is mass of the hypothetical terrestrial exoplanet, $e$ is eccentricity, $i_{\text{orb}}$ is the orbital inclination measured from the plane perpendicular to the sky, $\Omega$ is the longitude of ascending node, $\omega$ is the argument of pericenter, and $M$ is the mean anomaly. Since our timing precisions are at the level of a signal from a terrestrial planet, we limited the mass coverage to the ``super-Earth'' range of 1 -- 10 \mearth\ planets. Our semi-major axis coverage spans the 2:1 resonance, with one choice directly at the commensurability. All other parameters were varied so as to fully cover all possible architectures, but still to keep the entire program tractable. Each simulation required $\gtrsim 1$ hour of CPU time, or $\sim 7$ million CPU hours total.

\begin{table}
\begin{center}
\caption{\label{table_stability_XO2} Range of Parameter Space for Stability Calculations}
\begin{tabular}{cccccc}
\hline \hline
Parameter & Unit & Min & Max & $\Delta$ & $N$\\
\hline \hline
Mass & \mearth & 1 & 10 & 1 & 10\\
$a$ & AU & 0.055 & 0.062 & 0.0016 & 5\\
$e$ & - & 0 & 0.8 & 0.1 & 9\\
$i_{\text{orb}}$ & $^\circ$ & 0.001 & 70 & 10 & 8\\
$\Omega$ & $^\circ$ & 0 & 324 & 36 & 10\\
$\omega$ & $^\circ$ & 0 & 324 & 36 & 10\\
$M$ & $^\circ$ & 0 & 324 & 36 & 10\\
\hline
\end{tabular}
\end{center}
\end{table}

For the N-body simulations, we used the symplectic integrator code in the HNBody package \citep{rauchhamilton02}\footnote{Publicly available at http://janus.astro.umd.edu/HNBody/}, which includes general relativity. We integrated each trial for $10^6$ orbits of \xob\ ($\sim$ 7200 years), which is enough to identify $> 95$\% of unstable orbits for other exoplanet system \citep{barnesquinn04}. Each simulation could produce one of three outcomes: stable, unstable, or fail. Stable configurations lost no planets due to gravitational perturbations, while unstable ones did. Failed systems did not conserve energy to better than 1 part in $10^4$, which is required for symplectic integrators \citep{barnesquinn04}.

\begin{figure}[!h] 
\begin{center} 
\includegraphics[width=0.95\textwidth]{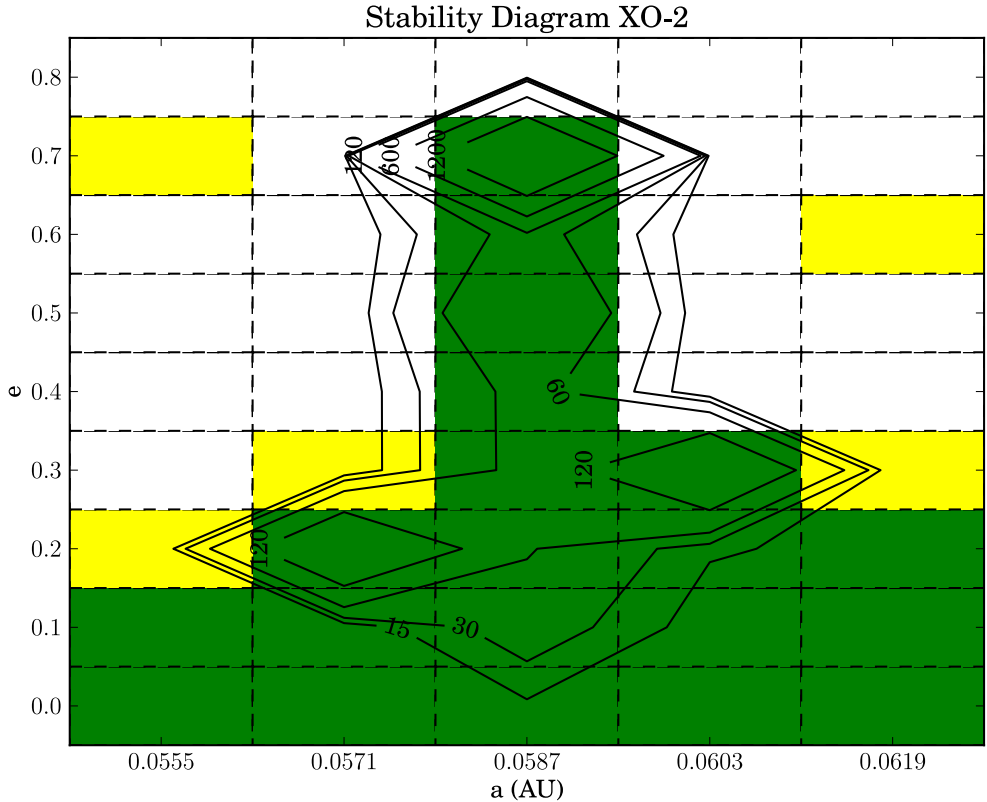}
\caption{\label{figure_stability_XO2}Contours of TTVs for orbital configurations of a 1\mearth\ companion to \xob, close to the 2:1 resonance. The color-coded squares note the stability of N-body simulations, with yellow (light grey) being unstable, white being failed, and green (dark grey) being stable simulations. The color version of this image is available in the online version.}
\end{center} 
\end{figure}

In Figure \ref{figure_stability_XO2}, a sample slice through the data set is shown. We limited the visualization to cases with $m \sim 1$ \mearth\ planets with $i_{\text{orb}} = \Omega = \omega = 0$, and $M = 144^\circ$. Green squares designate stable configurations, red unstable, and yellow failed. Most simulations from the yellow bins are probably unstable, as close approaches between planets can violate the algorithm's underlying assumption that the gravitational force from the star is much larger than that from any other planet. We note that the average fractional change in energy ($dE/E$) for stable cases was $7 \times 10^{-9}$.

The trials shown in Fig. \ref{figure_stability_XO2} show the that, for this value of $M$, the mean motion resonance at 0.0587~AU stabilizes the system. At all values of $a$ and low $e$, the system is stable, as predicted by Hill stability theory \citep[see][]{marchalbozis82, gladman93, barnesgreenberg06, kopparapubarnes10}. However at the resonance, stability is likely for $e \le 0.7$. As expected, we find that only certain values of $M$ predict this ``tongue'' of stability.

The contour lines in Fig. \ref{figure_stability_XO2} show the values of the TTV signal in seconds. For the broad stable region at $e \lesssim 0.1$, the signal is at or below the detection limit. For larger $e$, especially in the resonance, the value can be much larger. For the bin at $a =$ 0.0587~AU, and $e =$ 0.7, the signal magnitude is close to 2377~sec, or nearly 40 minutes. Clearly, the \apostle\ project could have detected an Earth-mass companion if it were in a favorable orbit. We do note, however, that an Earth-like planet with large eccentricity is likely to be rapidly tidally circularized due to the close proximity \citep{rasio96, jackson08, barnes10}, although some eccentricity would be maintained by the resonance.

As noted in Table \ref{table_OC_XO2} ($\S$ \ref{sec_ttv}) the median TTV errors obtained from APOSTLE lightcurves of \xob\ were $\sim$20~sec. Of the 3.6 million parameter configurations, $\sim$ 1.14 million were found to be stable, and approximately 51\% of these stable configurations produce TTVs on the order of 20~sec or less and thus would not be easily detected by APOSTLE or other surveys with similar capability. The vast amount of data generated by the stability simulations cannot be adequetly discussed in this text, hence we provide the results from all 3.6 Million simulations as part of an online-only table; a segment of which is displayed in Table \ref{table_simulation_XO2}. The first column `Sim. ID' denotes the simulation number and columns numbered (2)--(8) are the input configurations for a given simulation. The range of these simulation configurations were summarized earilier in this section and in Table \ref{table_stability_XO2}. Columns (9)--(13) are the simulation statistics. Column (9) indicates the stability outcome of a given simulation, with 0 or 1 denoting unstable and stable simulations respectively. Column (10) lists the energy conservation of a given simulaions as noted by the maximum fractional change in energy over a given step ($dE/E$). Columns (11), (12) and (13) list the standard deviations in the transit timing variations, transit duration variations and the variations in the impact parameter ($b$) respectively.

\begin{sidewaystable}
\begin{center}
\caption{\label{table_simulation_XO2} Data on the Stability Simulations of the XO-2 system}
\begin{tabular}{ccccccccccccc}
\hline \hline 
Sim. ID & Mass & $a$ & $e$ & $i$ & $\Omega$ & $\omega$ & $M$ & Stability & Energy Cons. & std(TTV) & std(TDV) & std(TBV) \\
 - & \mearth\ & AU & - & $^\circ$ & $^\circ$ & $^\circ$ & $^\circ$ & - & - & seconds & days & - \\
 (1) & (2) & (3) & (4) & (5) & (6) & (7) & (8) & (9) & (10) & (11) & (12) & (13) \\
 \hline \hline 

0&1.0&0.0555&0.0&0&0.0&0&0&1&2.370e-09&2.040e+00&2.656e-06&2.066e-08 \\
 1&1.0&0.0555&0.0&0&0.0&0&36&1&2.630e-09&2.062e+00&2.633e-06&2.000e-08 \\
 2&1.0&0.0555&0.0&0&0.0&0&72&1&2.610e-09&2.077e+00&2.710e-06&2.179e-08 \\
 3&1.0&0.0555&0.0&0&0.0&0&108&1&2.050e-09&2.068e+00&2.639e-06&2.165e-08 \\
 4&1.0&0.0555&0.0&0&0.0&0&144&1&1.980e-09&2.027e+00&2.684e-06&2.102e-08 \\
 5&1.0&0.0555&0.0&0&0.0&0&180&1&9.120e-10&2.029e+00&2.662e-06&1.983e-08 \\
 6&1.0&0.0555&0.0&0&0.0&0&216&1&2.210e-09&2.017e+00&2.680e-06&2.031e-08 \\
 7&1.0&0.0555&0.0&0&0.0&0&252&1&2.300e-09&2.050e+00&2.677e-06&2.140e-08 \\
 8&1.0&0.0555&0.0&0&0.0&0&288&1&2.430e-09&2.051e+00&2.673e-06&2.009e-08 \\
 9&1.0&0.0555&0.0&0&0.0&0&324&1&2.610e-09&2.031e+00&2.665e-06&1.958e-08 \\
 . &. &. &. &. &. &. &. &. &. &. &. &. \\
. &. &. &. &. &. &. &. &. &. &. &. &. \\
. &. &. &. &. &. &. &. &. &. &. &. &. \\
\hline
\end{tabular}
\footnotesize \begin{tabular}{l} 
*The data are presented in their entirety as an online-only table. \\
(1) Simulation ID, (2) Mass of perturber, (3) Semi-major axis, (4) Eccentricity, (5) Inclination, (6) Longitude of the ascendening node, \\
(7) the Argument of Pericenter, (8) the Mean Anomaly, (9) Stability Outcome, (10) Energy Conservation $dE/E$, \\
(11) Standard Deviation of TTV - Transit Timing Variations \\
(12) Standard Deviation of TDV - Transit Duration Variations \\
(13) Standard Deviation of TBV - Transit Impact Parameter (b) Variations \\
\end{tabular}
\end{center}
\end{sidewaystable}

\section{Conclusions}
\label{sec_conclusions}

\vspace{\baselineskip} \N \textbf{Photometric Precision:}
The \xo\ system was observed over a variety of observing conditions over a period of 3 years between 2008 -- 2011. The best photometric precisions we obtained with our \Iband\ and \rband\ observations were 405 and 354 ppm respectively (see Table \ref{table_ObsSum_XO2}).

\vspace{\baselineskip}  \N \textbf{\xob\ System Parameters:}
Our analysis of the 10 transit lightcurves yielded estimates of system parameters that agree with measurements presented by other studies (see Figure \ref{figure_MCMC_FLDC_DER_XO2}). We were able to improve the constraints on $a/R_{\star}$ and $\rho_{\star}$ by $\sim$40\% compared to the previous best measurements from the ground \citep{sing11}. The measurements are presented in Table \ref{table_ParTable03_XO2}; see the TAP values. We could not get the Markov chains to converge while fitting for stellar limb-darkening parameters, echoing the results from previous studies.

\vspace{\baselineskip}  \N \textbf{Search for Transit Depth Variations:}
Our Multi-Depth fits show some variations in the transit depth over transit epoch (see Figure \ref{figure_TDepthV_XO2} and Table \ref{table_ParTable02_XO2}) for the 6 \Iband\ lightcurves. Since 3 out of the 6 lightcurves were not fully sampled we cannot confidently assert real variability in the data. The seen variations could be shortcomings of the transit model's ability to fit a set of lightcurves with both complete and incomplete data. We do not see similar variations in the \rband, where one may expect spot-modulated variability to appear, due to the greater spot-to-star contrast in the \rband. There are no known reports of stellar activity on \xo, hence the \rband\ results are consistent with this fact.

\vspace{\baselineskip}  \N \textbf{Search for Transit Timing Variations:}
The \xob\ dataset contains lightcurves with some of the best photometric precisions achieved with ground-based observations. Since photometric precision directly translates to transit timing precision we are able to report timing precisions as low as 12~sec (after red-noise analysis, see Table \ref{table_OC_XO2}). We were unable to detect significant timing deviations for \xob\ in our data. The linear fit was not robust, with $\chi^2/DOF$ being significantly greater than 1, indicating large scatter around the linear ephemeris fit. The transit times derived from the \apostle\ lightcurves using the red-noise analysis of \citet{carterwinn09} resulted in more conservative errors than those derived using \tmcmc. A linear ephemeris is consistent with the transit time measurements reported from the TAP analysis. The overall variation in the O-C values from our rednoise analysis was on the order of $\sim 39$ sec. From a sinusoidal fit to \apostle's O-C data, we obtained a TTV amplitude of $\sim 27.4$ sec. However, checking the goodness-of-fit of sinusoids to random rearrangements of the data show that the detected periodicity unlikely to be real. The resulting amplitude rules out planetary companions more massive than 0.2 \mearth\ near the 2:1 MMR, and larger companions near higher order resonances.

We conclude that the set of transit times published in the literature for \xob\ and other transiting systems in general, are not suited for transit timing analysis. Lacking red-noise analysis leads to underestimated timing errors and may lead to premature reporting of timing variations. A proper analysis of transit times would need a simultaneous analysis of transit lightcurves using a transit model which is, (1) suited for Bayesian inference \citep[\ie\ with a fairly uncorrelated parameter set,][]{carter08} and (2) a transit model which can adequately account for red-noise in the data \citep[like TAP,][]{gazak11}. Using large data sets of transit lightcurves may be inefficient due to the slowness of Markov chains with the addition of model parameters. In addition to the development of more detailed models the utilization of fast Markov chain algorithms \citep[for e.g.][]{foreman-mackey12} is also recommended.

Our lack of detection of TTVs in the data for the Hot-Jupiter \xob\ is also consistent with \keptel's findings that (1) Hot-Jupiters tend to lack other planetary siblings \citep{latham11,steffen12} and (2) members of multi-planet systems with short period planets (Period $<$ 10 days) are more likely to be Hot-Neptunes \citep{latham11,lissauer11b}.

\vspace{\baselineskip}  \N \textbf{Dynamical Study:}
We ran 3.6 million N-body simulations of possible multi-planet configurations near the 2:1 resonance in order to test for (1) orbital stability and (2) detectability of TTV signals. We varied several properties of the hypothetical companion to \xob\ (see Table \ref{table_stability_XO2}), to look for patterns in the resulting stability and transit times of the simulations over parameter space. Of the several stable configurations we find that $\sim$51\% of the simulations would display TTVs weaker than the precision limits of our survey and hence remain undetectable. The entire table of simulation statistics are presented online (see Table \ref{table_simulation_XO2}).

\section*{Acknowledgments}
Funding for this work came from NASA Origins grant NNX09AB32G and NSF Career grant 0645416. RB acknowledges funding from NASA Astrobiology Institute's Virtual Planetary Laboratory lead team, supported by NASA under cooperative agreement No. NNH05ZDA001C. Data presented in this work are based on observations obtained with the Apache Point Observatory 3.5-meter telescope, which is owned and operated by the Astrophysical Research Consortium. We would also like to thank an anonymous referee for giving several helpful suggestions to improve our paper. We would like to thank the APO Staff, APO Engineers, Anjum Mukadam and Russel Owen for helping the APOSTLE program with its observations and instrument characterization. We would like thank S. L. Hawley for scheduling our observations on APO. This work acknowledges the use of parts of J. Eastman's EXOFAST transit code as part of APOSTLE's transit model \mtq. We also acknowledge the use of the \texttt{PyAstronomy} package\footnote{http://www.hs.uni-hamburg.de/DE/Ins/Per/Czesla/PyA/PyA/index.html}.


\begin{thebibliography}{}
\bibitem[Agol et al.(2005)]{agol05} Agol, E., Steffen, J., Sari, R., \& Clarkson, W.\ 2005, \mnras, 359, 567
\bibitem[Ballard et al.(2011)]{ballard11} Ballard, S., Fabrycky, D., Fressin, F., et al.\ 2011, \apj, 743, 200
\bibitem[Barnes \& Greenberg(2006)]{barnesgreenberg06} Barnes, R., \& Greenberg, R.\ 2006, \apjl, 647, L163
\bibitem[Barnes \& Quinn(2004)]{barnesquinn04} Barnes, R., \& Quinn, T.\ 2004, \apj, 611, 494
\bibitem[Barnes et al.(2010)]{barnes10} Barnes, R., Raymond, S.~N., Greenberg, R., Jackson, B., \& Kaib, N.~A.\ 2010, \apjl, 709, L95
\bibitem[Becker et al.(2013)]{becker13} Becker, A.~C., Kundurthy, P., Agol, E., et al.\ 2013, \apjl, 764, L17
\bibitem[Bertin \& Arnouts(1996)]{bertinarnouts96} Bertin, E., \& Arnouts, S.\ 1996, \aaps, 117, 393
\bibitem[Bessell(1990)]{bessell90} Bessell, M.~S.\ 1990, \pasp, 102, 1181
\bibitem[Brown et al.(2001)]{brown01} Brown, T.~M., Charbonneau, D., Gilliland, R.~L., Noyes, R.~W., \& Burrows, A.\ 2001, \apj, 552, 699
\bibitem[Burke et al.(2007)]{burke07} Burke, C.~J., McCullough, P.~R., Valenti, J.~A., et al.\ 2007, \apj, 671, 2115
\bibitem[Burrows et al.(2007)]{burrows07} Burrows, A., Hubeny, I., Budaj, J., Knutson, H.~A., \& Charbonneau, D.\ 2007, \apjl, 668, L171
\bibitem[Carter et al.(2008)]{carter08} Carter, J.~A., Yee, J.~C., Eastman, J., Gaudi, B.~S., \& Winn, J.~N.\ 2008, \apj, 689, 499
\bibitem[Carter \& Winn(2009)]{carterwinn09} Carter, J.~A., \& Winn, J.~N.\ 2009, \apj, 704, 51
\bibitem[Charbonneau et al.(2000)]{charbonneau00} Charbonneau, D., Brown, T.~M., Latham, D.~W., \& Mayor, M.\ 2000, \apjl, 529, L45
\bibitem[Claret \& Bloemen(2011)]{claretbloemen11} Claret, A., \& Bloemen, S.\ 2011, \aap, 529, A75
\bibitem[Crouzet et al.(2012)]{crouzet12} Crouzet, N., McCullough, P.~R., Burke, C., \& Long, D.\ 2012, \apj, 761, 7
\bibitem[Collier Cameron et al.(2007)]{colliercameron07b} Collier Cameron, A., et al.\ 2007, \mnras, 380, 1230
\bibitem[Cousins(1976)]{cousins76} Cousins, A.~W.~J.\ 1976, Monthly Notes of the Astronomical Society of South Africa, 35, 70
\bibitem[Eastman et al.(2010)]{eastman10} Eastman, J., Siverd, R., \& Gaudi, B.~S.\ 2010, \pasp, 122, 935
\bibitem[Eastman et al.(2012)]{eastman12} Eastman, J. et al., in prep
\bibitem[Fernandez et al.(2009)]{fernandez09} Fernandez, J.~M., Holman, M.~J., Winn, J.~N., et al.\ 2009, \aj, 137, 4911
\bibitem[Ford(2005)]{ford05} Ford, E.~B.\ 2005, \aj, 129, 1706
\bibitem[Foreman-Mackey et al.(2012)]{foreman-mackey12} Foreman-Mackey, D., Hogg, D.~W., Lang, D., \& Goodman, J.\ 2012, arXiv:1202.3665
\bibitem[Fortney et al.(2008)]{fortney08} Fortney, J.~J., Lodders, K., Marley, M.~S., \& Freedman, R.~S.\ 2008, \apj, 678, 1419
\bibitem[Fukugita et al.(1996)]{fukugita96} Fukugita, M., Ichikawa, T., Gunn, J.~E., Doi, M., Shimasaku, K.,\& Scheider, D.~p.\ 1996, \aj, 111, 1748
\bibitem[Gazak et al.(2011)]{gazak11} Gazak, J.~Z., Johnson, J.~A., Tonry, J., et al.\ 2011, arXiv:1102.1036
\bibitem[Gelman \& Rubin(1992)]{gelmanrubin92} Gelman, A. \& D. B. Rubin, \ 1992, Statistical Science, 7, 457
\bibitem[Gelman et al.(2003)]{gelman03} Gelman, A. et al. \ 2003, Bayesian Data Analysis (2nd ed.; Boca Raton: Chapman \& Hall/CRC)
\bibitem[Gladman(1993)]{gladman93} Gladman, B.\ 1993, \icarus, 106, 247
\bibitem[Haghighipour \& Kirste(2011)]{haghighipourkirste11} Haghighipour, N., \& Kirste, S.\ 2011, Celestial Mechanics and Dynamical Astronomy, 111, 267
\bibitem[Holman \& Murray(2005)]{holmanmurray05} Holman, M.~J., \& Murray, N.~W.\ 2005, Science, 307, 1288
\bibitem[Holman et al.(2010)]{holman10} Holman, M.~J., Fabrycky, D.~C., Ragozzine, D., et al.\ 2010, Science, 330, 51
\bibitem[Hubeny et al.(2003)]{hubeny03} Hubeny, I., Burrows, A., \& Sudarsky, D.\ 2003, \apj, 594, 1011
\bibitem[Jackson et al.(2008)]{jackson08} Jackson, B., Greenberg, R., \& Barnes, R.\ 2008, \apj, 678, 1396
\bibitem[Knutson et al.(2007)]{knutson07} Knutson, H.~A., Charbonneau, D., Noyes, R.~W., Brown, T.~M., \& Gilliland, R.~L.\ 2007, \apj, 655, 564
\bibitem[Knutson et al.(2010)]{knutson10} Knutson, H.~A., Howard, A.~W., \& Isaacson, H.\ 2010, \apj, 720, 1569
\bibitem[Kopparapu \& Barnes(2010)]{kopparapubarnes10} Kopparapu, R.~K., \& Barnes, R.\ 2010, \apj, 716, 1336
\bibitem[Kundurthy et al.(2011)]{kundurthy11} Kundurthy, P., Agol, E., Becker, A.~C., et al.\ 2011, \apj, 731, 123
\bibitem[Kundurthy et al.(2013)]{kundurthy13} Kundurthy, P., Becker, A.~C., Agol, E., Barnes, R., \& Williams, B.\ 2013, \apj, 764, 8
\bibitem[Latham et al.(2011)]{latham11} Latham, D.~W., Rowe, J.~F., Quinn, S.~N., et al.\ 2011, \apjl, 732, L24
\bibitem[Lissauer et al.(2011a)]{lissauer11a} Lissauer, J.~J., Fabrycky, D.~C., Ford, E.~B., et al.\ 2011, \nat, 470, 53 
\bibitem[Lissauer et al.(2011b)]{lissauer11b} Lissauer, J.~J., Ragozzine, D., Fabrycky, D.~C., et al.\ 2011, \apjs, 197, 8
\bibitem[Machalek et al.(2008)]{machalek08} Machalek, P., McCullough, P.~R., Burke, C.~J., et al.\ 2008, \apj, 684, 1427
\bibitem[Machalek et al.(2009)]{machalek09} Machalek, P., McCullough, P.~R., Burrows, A., et al.\ 2009, \apj, 701, 514
\bibitem[Madhusudhan(2012)]{madhusudhan12} Madhusudhan, N.\ 2012, \apj, 758, 36 
\bibitem[Mandel \& Agol(2002)]{mandelagol02} Mandel, K., \& Agol, E.\ 2002, \apjl, 580, L171
\bibitem[Marchal \& Bozis(1982)]{marchalbozis82} Marchal, C., \& Bozis, G.\ 1982, Celestial Mechanics, 26, 311
\bibitem[Monet et al.(2003)]{monet03} Monet, D.~G., Levine, S.~E., Canzian, B., et al.\ 2003, \aj, 125, 984
\bibitem[Mukadam et al.(2011)]{mukadam12} Mukadam, A.~S., Owen, R., Mannery, E., et al.\ 2011, \pasp, 123, 1423
\bibitem[Nesvorn{\'y} et al.(2012)]{nesvorny12} Nesvorn{\'y}, D., Kipping, D.~M., Buchhave, L.~A., et al.\ 2012, Science, 336, 1133
\bibitem[Rasio et al.(1996)]{rasio96} Rasio, F.~A., Tout, C.~A., Lubow, S.~H., \& Livio, M.\ 1996, \apj, 470, 1187
\bibitem[Rauch \& Hamilton(2002)]{rauchhamilton02} Rauch, K.~P., \& Hamilton, D.~P.\ 2002, Bulletin of the American Astronomical Society, 34, 938
\bibitem[Roberts \& Rosenthal(2009)]{robertsrosenthal09} Roberts, G. O. \& Rosenthal, J. S.\ 2009, Journal of Computational \& Graphical Statistics, 18, 349
\bibitem[Sing et al.(2011)]{sing11} Sing, D.~K., D{\'e}sert, J.-M., Fortney, J.~J., et al.\ 2011, \aap, 527, A73
\bibitem[Southworth(2010)]{southworth10} Southworth, J.\ 2010, \mnras, 408, 1689
\bibitem[Steffen et al.(2012)]{steffen12} Steffen, J.~H., Ragozzine, D., Fabrycky, D.~C., et al.\ 2012, arXiv:1205.2309
\bibitem[Tegmark et al.(2004)]{tegmark04} Tegmark, M., et al.\ 2004, \prd, 69, 103501
\bibitem[Winn(2011)]{winn11} Winn, J.~N.\ 2011, Exoplanets, edited by S.~Seager.~ Tucson, AZ: University of Arizona Press, 2011, 526 pp.~ ISBN 978-0-8165-2945-2., p.55-77, 55
\bibitem[Zechmeister \& K\"{u}rster(2009)]{zechmeisterkurster09} Zechmeister, M., K\"{u}rster, M.\ 2009, \aap, 496, 577
\end{thebibliography}
\end{document}